\documentclass{phb-proc4-auth}

\usepackage{graphicx}
\usepackage{amssymb}

\begin{document}
\begin{frontmatter}


\journal{SCES '04}


\title{Crossover from Single-Ion to Coherent Non-Fermi Liquid Behavior
in Ce${}_{1-x}$La${}_x$Ni${}_9$Ge${}_4$}

%
%
%
%
%
%

\author[CPM]{E. - W. Scheidt  \thanksref{SFB}}
\author[CPM]{U. Killer }
\author[Wien]{H. Michor}
\author[Wien]{E. Bauer}
\author[Wien]{C. Dusek}
\author[TP3]{S. Kehrein \thanksref{SFB}\thanksref{HEI}}
\author[CPM]{W. Scherer \thanksref{SFB}}

%

\address[CPM]{Chemische Physik und Materialwissenschaften,
Universit\"{a}t Augsburg, 86159 Augsburg, Germany}
\address[Wien]{Institut f\"ur Festk\"orperphysik, TU Wien, 1040 Wien,
Austria}
\address[TP3]{TP III, Elektronische Korrelationen und Magnetismus,
Universit\"{a}t Augsburg, 86159 Augsburg, Germany}

%
%
%
%

\thanks[SFB]{This work was supported by the SFB~484 of the Deutsche
Forschungsgemeinschaft (DFG).}
\thanks[HEI]{The author acknowledges support through a
Heisenberg fellowship of the DFG.}

%
%
%
%



\begin{abstract}

We report specific heat and magneto-resistance studies on the
compound Ce${}_{1-x}$La${}_x$Ni${}_9$Ge${}_4$ for various
concentrations over the entire stoichiometric range. Our data
reveal single-ion scaling with Ce-concentration between $x = 0.1$
and $0.95$. Furthermore, CeNi${}_9$Ge${}_4$ turns out to have the
largest ever recorded value of the electronic specific heat
$\Delta c/T \approx$ 5.5\,J\,$\rm K^{-2}mol^{-1}$ at $T=0.08$\,K
which was found in Cerium f-electron lattice systems. In the doped
samples $\Delta c/T$ increases logarithmically in the temperature
range between $3$\,K and $50$\,mK typical for non-Fermi liquid
(nFl) behavior, while $\rho$ exhibits a Kondo-like minimum around
30\,K, followed by a single-ion  local nFl behavior. In contrast
to this, CeNi${}_9$Ge${}_4$ flattens out in $\Delta c/T$ below
$300$\,mK and displays a pronounced maximum in the resistivity
curve at 1.5\,K indicating a coherent heavy fermion groundstate.
These properties render the compound
Ce${}_{1-x}$La${}_x$Ni${}_9$Ge${}_4$ a unique system on the
borderline between Fermi liquid and nFl physics.

\end{abstract}

%
%

\begin{keyword}

non Fermi liquid \sep Heavy Fermion \sep single ion effect

\end{keyword}


\end{frontmatter}

%
%
%
%
%

Cerium-based intermetallic compounds exhibit a rich variety of
ground state properties due to the competition between three
different types of interactions: i) crystal-field effects; ii)
Ce-Ce intersite correlations which result in long-range magnetic
order and iii) on-site correlation between 4f- and
conduction-electron states leading to a formation of a local Kondo
singlet. In order to minimize the intersite correlation by
increasing the Ce-sublattice space CeNi${}_9$Si${}_4$ and
CeNi${}_9$Ge${}_4$ turn out to be  good candidates. Both compounds
crystallize in a tetragonal structure with space group
\emph{I}4/\emph{mcm} and a minimum Ce-Ce distance of
$5.5479\,{\rm\AA}$ for Si \cite{Michor_03} and of
$5.6357\,{\rm\AA}$ for Ge \cite{Michor_04}. While
CeNi${}_9$Si${}_4$ is a typical Kondo lattice system with a
moderate Sommefeld coefficient $\gamma = 155$\,mJ\,$\rm
K^{-2}mol^{-1}$ \cite{Michor_03}, CeNi${}_9$Ge${}_4$ exhibits
non-Fermi liquid behavior with the largest ever recorded value of
the electronic specific heat $\Delta c/T \approx$ 5.5\,J\,$\rm
K^{-2}mol^{-1}$ at $T=0.08$\,K without showing any trace of
magnetic order \cite{Michor_04}.
\begin{figure}
\centering
\includegraphics[width=7cm,clip]{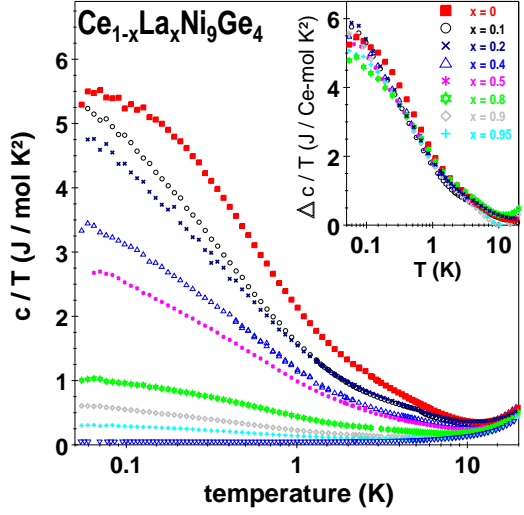}
\caption{A semi-logarithmic plot of the specific heat divided by
temperature of Ce${}_{1-x}$La${}_x$Ni${}_9$Ge${}_4$ in the
temperature range $0.05 K < T < 20 K$.The insert shows the
electronic contribution to the specific heat $\Delta$c/T
normalized per Ce-mol.}
\label{fig1}                          
\end{figure}
Recently, specific heat and susceptibility studies on
La-substituted samples Ce${}_{1-x}$La${}_x$Ni${}_9$Ge${}_4$
indicate that the large electronic specific heat value $\Delta c/T
$ of CeNi${}_9$Ge${}_4$ is mainly caused by single ion effects
\cite{Killer_04}. In addition the observed nFl-behavior in $\Delta
c/T$ and $\rho$ and the Kondo-like behavior in $\chi$ suggests
that in the Ce${}_{1-x}$La${}_x$Ni${}_9$Ge${}_4$ system two
channel Kondo physics takes place \cite{Killer_04}.
\begin{figure}
\centering
\includegraphics[width=6.8cm,clip]{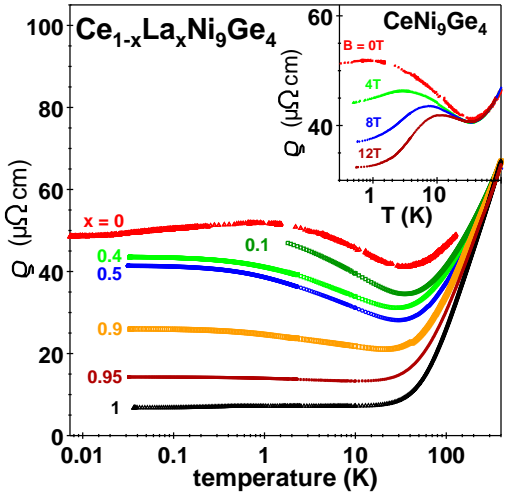}
\caption{The electrical resistivity $\rho(T)$ of various samples
normalized at 300\,K to that of LaNi${}_9$Ge${}_4$ which was
detected with the Vander-Pauw method. Notice that the unexpected
smaller residual resistivity ratio of the stochiometric single
crystal sample \cite{Bayreuth} may be due to anisotropy effects
and stress. The insert shows for CeNi${}_9$Ge${}_4$ the shift of
the resistivity maximum at 1.5\,K with increasing magnetic field
up to higher temperatures.}
\label{fig2}                          
\end{figure}

The specific heat divided by temperature  of polycrystalline
samples Ce${}_{1-x}$La${}_x$Ni${}_9$Ge${}_4$(with $x$ ranging from
0 to 1) is displayed in Fig.~\ref{fig1}. For all La substituted
samples a nearly logarithmic increase of  $ c/T$ below 1.5\,K is
observed which is characteristic for nFl physics. Only the
non-diluted compound CeNi${}_9$Ge${}_4$ deviates noticeably from
this logarithmic behavior indicating collective excitations below
0.3\,K \cite{Michor_04}. Furthermore, the electronic contribution
normalized per Ce-concentration is found to be almost independent
of La substitution (insert Fig.~\ref{fig1}). This strongly
suggests that the low-temperature physics in
Ce$_{1-x}$La$_{x}$Ni$_{9}$Ge$_{4}$ is governed by single-ion
behavior, at least for $x>0$.

The concentration dependent crossover from single-ion nFl to
coherent behavior in Ce${}_{1-x}$La${}_x$Ni${}_9$Ge${}_4$ is also
supported by resistivity measurements (Fig.~\ref{fig2}). With the
exception of CeNi${}_9$Ge${}_4$ and LaNi${}_9$Ge${}_4$ the
resistivity of all samples passes through a Kondo-like minimum
around 30\,K, followed by a logarithmic temperature increase and
saturates below 2\,K with $\rho(T)-\rho(0) \propto T^{c}$, c =
0.8$\pm$0.2 indicating single-ion nFl-like behavior. While
LaNi${}_9$Ge${}_4$ shows normal metallic behavior, only
CeNi${}_9$Ge${}_4$ has a maximum at low temperature indicating a
concentration dependent crossover from a single-ion to a coherent
Kondo lattice state below 1\,K.
\begin{figure}
\centering
\includegraphics[width=7cm,clip]{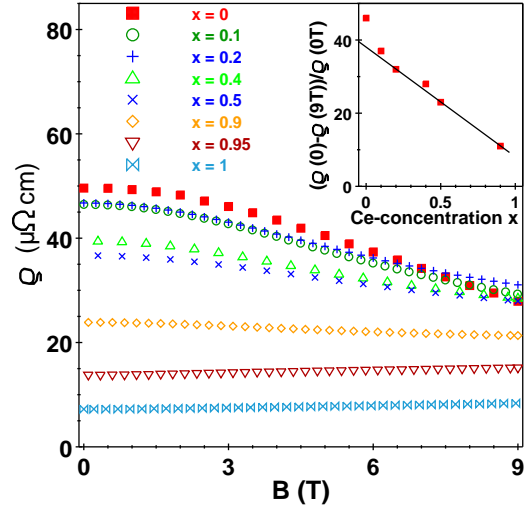}
\caption{Magneto resistance of
Ce${}_{1-x}$La${}_x$Ni${}_9$Ge${}_4$ at 2\,K. Insert: percentage
decrease of magneto resistance between 0 and 9T.}
\label{fig3}                          
\end{figure}
Fig.~\ref{fig3} shows the magneto-resistance between 0 and 9 T at
2K. The percentage decrease of magneto resistance between
$\rho_{imp}(0)$ and $\rho_{imp}(9T)$ increases linearly with
rising Ce-concentration starting at x = 0.9 (insert
Fig.~\ref{fig3}). Only for the non-diluted CeNi${}_9$Ge${}_4$ this
reduction deviates from the linearity to higher values, again
indicating a concentration dependent crossover from single-ion nFl
to coherent behavior. Therefore the magneto-resistance data are in
good agreement with those of the specific heat data
\cite{Killer_04}.

%
%
%
%

%
%
%
%


\end{document}